# Hybrid Structured Editing

## Structures for Tools, Text for Users

Tom Beckmann[a], Christoph Thiede[a], Jens Lincke[a], and Robert Hirschfeld[a]

a   Hasso Plattner Institute, University of Potsdam, Germany

**Abstract**   In programming, better tools often yield better results. For that, modern programming environments offer mechanisms to allow for their extensibility. The closer those tools are to the code, the easier it is for programmers to map the information provided by a tool to the code this information is about.

However, existing extension mechanisms do not facilitate the close integration of tools with textual source code. Tools must be able to track program structures across edits to appear at the right positions but the parsing step of text complicates tracking structures.

We propose *hybrid structured editing*, an approach that supports tool builders by providing structural guarantees while providing tool users with a familiar and consistent text editing interface.

*Hybrid structured editing* allows tool builders to declare constraints on the structure that a program must conform to and ensures their observance.

We present an implementation and several case studies of tools based on *hybrid structured editing* to demonstrate its effectiveness.

*Hybrid structured editing* supports the safe extension of programming environments with tools that work on a structured representation of code and provide a consistent and reliable user experience.

**ACM CCS 2012**

- **Software and its engineering** → Visual languages; *Domain specific languages*; **Integrated and visual development environments**;

**Keywords**   structured editing, programming tools

## The Art, Science, and Engineering of Programming

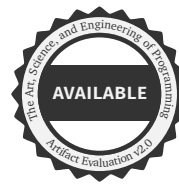
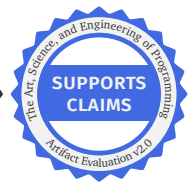

**Hybrid Structured Editing**

```
1  function processList(list) {
2      Example: ascending  Self: null  Args: [[2, 3, 4]] ;
3
4    return  list.map(n => n ** 3) ;
                [8,27,64]
5  }
```

■ **Figure 1** A Babylonian Programming [39] example definition that executes the shown function and a watch showing the latest execution result.

## 1 Introduction

Integrated development environments (IDEs) feature large suites of tools that support programmers. These range from tooltips on source code elements that pull in documentation over integrated debuggers to visual designers for static user interface descriptions. These tools augment programmers' perspectives on their programs beyond what is visible from just the source code's text.

Most of these tools appear as lightweight decorations on the textual source code or in separate panels or windows. Often, tools could benefit from appearing closer to the specific parts of source code that they refer to, to minimize the spatial distance between related information and thus help the user understand their relationship [43].

A number of tools in research demonstrate the value of such tools that closely integrate with the text, as seen in Figure 1. For example, in Babylonian Programming [39] probes attach directly to expressions in source code to show the runtime values that these expressions assume. Interactive Visual Syntax [1] shows how nesting visual means to express domain-specific aspects of programs can make algorithms easier to understand. Code Bubbles [9] rearranges the parts of source code files to better communicate their relationships. We call these tools *code-centric tools*. Creating code-centric tools presents multiple difficult challenges, however. Among these, we focus on three major challenges:

**Structure Tracking** Tools must track structures of the program tree across edits to make sure that the tool appears at stable and expected locations in the source code, even when source code related to that structure changes.

**Nested Editors** Tools must be able to provide code editing facilities, even if these are deeply nested in the tool's user interface or concern just a small part of the original source text. Editing and navigating must remain predictable for the user, even in the presence of disjunct editors.

**Structured Access** Tools must be able to read static and dynamic information of the source code, map that information to relevant program structures, and must be able to edit the structures while maintaining correct syntax.

While, as described above, a number of tools demonstrate that addressing these challenges is possible, each tool implementation has to invest significant effort into addressing these three challenges, potentially driving tool builders to create tools with less optimal forms of display for their users that are simpler to implement.

As potential point of evidence, a number of code-centric tools published in research build on top of *structured editors* [5, 27, 37, 41], where working with structures and fragments of source code is significantly simpler for tools. In a structured editor, users do not edit text but edit the structures of the program tree directly. Consequently, the program tree is valid at all times and structures maintain their identity across edit operations. As a downside of structured editors, it is difficult to design edit operations that appear familiar to users coming from textual editing [3, 7, 44].

In this paper, we propose *hybrid structured editing* (HybridSE): a mechanism for extending normal text editors with *selective structural guarantees* and *fragments* for recursive nesting of tools and editors, which addresses the three challenges for code-centric tools. HybridSE forms a generalization of challenges encountered in prior implementation of Babylonian Programming and similar graphical tools that augment source code, which we discuss in Section 2. To demonstrate its feasibility, we describe implementations of HybridSE for the JavaScript code editor CodeMirror, lively4, and Squeak/Smalltalk and show a range of tool case studies to evaluate it. To integrate HybridSE with other editors, these editors must support embedding graphical elements within the textual code and implement a small interface described in Section 5, which allows HybridSE to take care of the lifetime and placement of the graphical elements. HybridSE is independent of languages and their paradigms—its only requirement is that the language is parsed to a program tree such that changes on a structural level can be determined.

In the following, we first describe the three challenges and related work further (Section 2), before giving an example of defining tools in HybridSE (Section 3). We then describe the concepts enabling HybridSE (Section 4), as well as our reference implementation (Section 5). Finally, we present a set of case studies (Section 6) and discuss our approach (Section 7) before concluding the paper (Section 8).

## 2 Background and Related Work

In this section, we further describe the three challenges we established in Section 1 and how related work addresses these.

### 2.1 Tracking Program Tree Structures

As described in Section 1, one challenge for code-centric tools is to track structures of the program tree. Take, for example, a tool such as Babylonian Programming: *probes* wrap around an expression and show live values observed from its runtime execution. To do so, probes must identify the correct range of text that the probed expression describes and must be able to track it across edits by the user.

A probe in a Babylonian Programming system is initially created through a shortcut: the user selects a text range and hits the probe shortcut. The system then accesses the current parse tree of the program to find the smallest expression that contains the user's selection in the source code text. This expression then receives the probe and the probe widget is shown attached to its text range.

**Hybrid Structured Editing**

When an edit occurs, the Babylonian Programming system now has to identify the new text range that the probed expression belongs to. A number of heuristics are available to accomplish this. As a start, the system may shift the text range by the number of added or removed characters if the edit happened before the probe's current text range. If the edit was within its text range, we instead extend or shrink the text range accordingly. However, when the edit overlaps the start or end point of the text range, this simple method may yield an invalid text range and we have to resort to re-parsing the file and selecting the closest expression that would correspond to our previous text range, adjusted by the change.

Notably, an edit can also introduce syntax errors. While syntax errors are present, it is no longer possible to parse the file correctly. As this challenge is also observed for other language support tooling, such as syntax highlight, research in parsers proposed various error recovery mechanisms to provide useful outputs in the presence of syntax errors [12]. However, even with error recovery, it is possible that it becomes intermittently unclear what text range our tracked expression corresponds to.

As an alternative, structured editors require users to express edit operations on the structure-level of the program tree directly, thus removing any ambiguity from textual editing. These structured editors thus serve as great basis for tool creation [8, 16, 17, 27, 40, 41]. A mismatch occurs, as these editors often seek compatibility or resemblance to text editors and their languages. Prior work has investigated ways to make editing text-like or text languages usable in structured editors [3, 31, 35, 44, 45]. Our approach in this paper, instead, seeks to make structures accessible to tools but leave text accessible to editor users.

In other text editors, only limited means to work with structures are available. For example, in the language server protocol [33], users identify structures by their location in the text buffer or from a snapshot of the current parsing state. Approaches for diffing of program trees allow identifying changes between two states of a program [19, 20]. In hybrid structured editing, we integrate program tree diffing to hide the complexity of manually diffing and tracking structures from tool builders.

**2.2 Nesting Editors**

Research projects such as Code Bubbles [9] demonstrate the possibilities programming tools can provide to programmers by rearranging source code to better suit their tasks. However, most IDEs support only a monolithic view on source code, which, at a maximum, shows a subset of lines of code in a separate view or the content of a notebook cell [38]. Examples in research show how subprograms can be composed, as for example with Eco [13], PolyJuS [36], Moonchild [14], or Engraft [22]. Instead, we aim to enable tools to handle fragments of code that are as small as a meaningful character, such as a single expression. When showing a source file in small fragments, various challenges arise. For example, navigation at the boundaries of fragments may have to jump to a visually adjacent fragment, jump to a fragment that shows an adjacent text location, or remain locked within the fragment.

Further, a fragment will likely display the text range of one specific node in the program tree. Consequently, management of fragments faces the same issues as probes

■ **Listing 1** Illustration of isolating a text range that spans multiple indented lines.

```
1  // original full source file of which we want to isolate the method argument
2  function func() {
3      if (…) {
4          method(array
5              .map()
6              .split())
7      }
8  }
9  // just the text range of the isolated method argument -- the indentation is kept
10 // from the original context but looks off when isolated
11 array
12         .map()
13         .split()
```

in Babylonian Programming as described in Section 2.1. When the user performs a change at the boundaries of a fragment, they will likely expect that the change will also become visible in that fragment but, due to tracking of a node's text range, it is possible that a partial change will not be considered part of that fragment. As an example, consider a number that is changed to an addition with a number.

```
1  2
2  2 + 3
```

When the user types a space character before typing the plus sign, the space character may appear to have been ignored, as the tracked expression, the number, still only encompasses the character 2.

Finally, when isolating an expression in a fragment that spans multiple lines, formatting can become an issue. For example, the expression may have been placed in a location with a large indentation. Displayed in the fragment, the indent then shifts all but the first line to the right, as illustrated in Listing 1. For example in Code Bubbles, this problem is addressed by automatically re-formatting each bubble with code given the local indentation level.

## 2.3 Structured Access to Program Trees

As the final challenge in Section 1, we established reading static and dynamic information from the program, as well as performing edits in a way that maintains correctness of the program's syntax. Reading static information is comparatively simple, as it only requires an interface to access and navigate structures of the parsed program tree. However, reading dynamic information in a way that is meaningful for a tool attached to a specific node of the program tree requires a mechanism that allows mapping information that is collected at program runtime to the specific node of interest. Beyond this, it may also require collecting contextual information, such as the current stack, to make effective use of the collected information.

For editing the program tree, a potentially surprising number of corner cases may arise, depending on the language grammar. For example, changing the content of a string requires observing escaping rules for quotation marks, newline characters, and other special characters. Or, nesting an expression as left-hand side of a binary operation requires inspecting its structure and the structure of its parent to find whether parentheses have to be placed to arrive at the intended meaning.

Looking at related work more generally, language workbenches [18] are designed to facilitate the creation of new languages—which can include making available information about the language's structures to tools. Another approach to make static information about programs available in editors is the language server protocol [33] (LSP). The LSP addresses the *IDE portability problem* [26], where every tool had to be adapted for one specific editor's extension mechanism. With the LSP, editors implement a single protocol, acting as *language clients*, and language support tooling also implements a single protocol, acting as *language servers*, such that a connection between arbitrary editors and language support servers is possible. Beyond these, some frameworks and systems are specifically designed to facilitate working with program structure information, such as Moldable Tools [11], Lorgnette [21], or Vivide [42]. To bridge from runtime execution back to the editor, source maps [25] are one approach that store a mapping from original source to generated sources.

## 3 Definition of Code-centric Tools

In this section, we introduce the framework that HybridSE enables for defining code-centric tools from the point of view of the tool builder. In Section 4, we then describe the underlying HybridSE system. The framework mirrors the flow of data as a sequence of five steps from **matching**, **extracting**, and **constraining** information of program structures, to defining **views** to show this information and defining **interactions** that cause changes to the program structures.

To illustrate the framework, we describe a watch tool that reports runtime information for an expression that the user has marked. Our watch tool should appear around all expressions that are wrapped in a special marker construct of the form:

```
1  ["__watch", list.map(n => n ** 3)][1]
```

The marker construct is an array that has the original expression as its second element. The marker's first array element is a special string that we can match against. The array expression evaluates to its second element, such that the marked expression can simply be used in place of the original expression.

Our goal is now to replace this marker expression with a watch user interface that shows the values the list.map(n => n ** 3) expression assumes during runtime and hides the marker construct, as shown in Figure 1. Including the marker in the source code allows us to more easily identify the location of the watched expression as the user performs changes. In the explanations below, we refer to the listing in Listing 2 that shows the tool definition by line numbers.

**(1) Program Structure Match** First, the tool builder defines a query that matches structures of the program. The query is a pure function that returns whether a given

■ **Listing 2** Declaration of the watch tool seen in Figure 1.

```
const watchTool = {
    query: node => {
        const match = node.matchesTemplate('["__watch", $expression][1]');
        return !match ? null : {
            expressionStructure: match.expression,
            expressionValues: match.expression.values,
        };
    },
    constraints: [tool => node.matchesTemplate('["__watch", $expression][1]')],
    view: ({ expressionStructure, expressionValues }) => {
        const removeWatch = () => {
            tool.node.replaceWith(expressionStructure, { intentDeleteNodes: [tool.node] });
        };
        return <div style="background: black; padding: 1rem">
            <Fragment nodes={[expressionStructure]} />
            <div style="color: white">{expressionValues.last()?.toString()}</div>
            <div class="close-icon" onclick={removeWatch}></div>
        </div>
    },
    type: "replace",
}
```

program structure conforms to the structure that the tool is looking for. HybridSE then runs this query function against all structures visible in editors and instantiates the tool for each match that occurs.

We are using a matching syntax from prior work [4], that allows specifying the literal syntax to be matched and variation points using a dollar-prefix element, instead of using, e.g., imperative access to the nodes' fields to find if they match,[1] as seen in Line 3. If a match failed, we return `null` in Line 4 to inform the system that our tool should not be instantiated for this program structure.

**(2) Extract Data** Next, the tool builder extracts information from the structures that were matched. The conceptual framework offers a navigation interface that allows tool builders to access children, parents, and data of structures. In addition, it offers a general mechanism to subscribe to observed runtime values that a structure assumes, if applicable.

For the above example of a watch, we extract both the expression structure itself in Line 5, as well as a stream that produces the runtime values observed for that expression during execution in Line 6.

**(3) Constraints** Given the extracted data from the program structures, the tool builder can now proceed to declare constraints that HybridSE will maintain. Constraints in our framework are pure functions that receive the new program tree and a set of changes to the program tree and return a boolean that indicates whether these changes would violate the constraint.

---

[1] If the dollar sign is valid language syntax, a preprocessing step can be added that replaces the character by some other sequence of characters that is valid but less likely to be used.

Since the watch in our example will contain state in its user interface, namely the history of past values, we want to ensure that a change does not intermittently make it disappear, which would be the case if the query we defined in Line 3 no longer matches. To prevent this, we ensure that our query still applies to the new tree. It is then HybridSE's task to ensure that this constraint is maintained before applying any edit to the program. We describe our HybridSE's response to a violation in Section 4.1.2.

**(4) View**   The means to define the view of the tool's user interface depends on the editing platform that the tool is hosted in. Our reference implementation, for example, is hosted in the browser and thus uses the browser's document object model to define its user interface.

To address the **Nested Editors** challenge formulated in Section 1, HybridSE requires host editors that implement HybridSE to offer *fragments*: these are full-featured code editors, meaning they support error highlights, autocompletion, and all other features of the host editor, but they show only the text range of one or multiple consecutive program structure nodes. Fragments can contain tools that, in turn, contain fragments, allowing for recursive nesting of editors. Navigation and selection between fragments and tools should be predictable from a user's perspective. Fragments also normalize the indentation level of the shown structures to address the issue described in Listing 1, for which we describe our approach in Section 4.2.

In our example, we define a user interface using HTML embedded in JavaScript, where values are interpolated using curly braces, as shown in Line 14. We embed a fragment in Line 15 that shows an editor for the watched expression and the last runtime value that was observed, if any. We also define that clicking the watch's container should remove the watch in Line 17.

Beyond views that **replace** the full matched structure as specified in Line 20, our framework also supports defining that a view should be **inserted before or after** the matched structure, or that **markup** should be added to the textual range corresponding to the text range corresponding to the structure. This markup could add a highlight, change the font size, or make the range clickable.

**(5) Interactions**   Lastly, we declare the callbacks that our view triggers in response to user interactions. For the watch, we thus declare the `removeWatch` function in Line 11. We use a program structure-aware programming interface for editing the tree, which we describe in more detail in Section 4.3.3. Here, we are using its `replaceWith` method that replaces the structure that the method is called on with the provided structure. In addition, we specify that it is the user's intent to delete the tool's node. Given this specification, HybridSE knows to ignore the constraint that requires the tool's node to remain in the tree.

**Result** This tool definition addresses the challenges established in Section 1:

**Structure Tracking** There is no explicit need to track structures across edits. HybridSE provides a stable reference to the node that the tool is instantiated on, and the constraint that the tool declared ensures that the node remains in the tree.

**Nested Editors** The fragment encapsulates the complexity of wrapping a user interface around a part of the source code.

**Structured Access** A structure-aware access and editing interface allows tool builders to work with structures at the level of abstraction that matches the formulation of their intention, both for static and dynamic information.

In the next section, we describe how HybridSE enables this conceptual framework.

## 4 Hybrid Structured Editing

HybridSE enables the framework described in Section 3. It is designed to be implemented in general-purpose programming editors. To allow a complete implementation of HybridSE, host editors need to provide the ability to:

1. integrate with navigation, undo, and editing events,
2. replace text ranges in the editor widget with a user interface widget, and
3. create instances of the editor widget that show a specific text range of a file, while editor functions continue to work in the context of the full file. HybridSE takes care of synchronizing the shown text range and text content but any other editor state (such as the mode for a modal editor integration like Vi) must be synchronized by the host.

An implementation of HybridSE in a host editor then is comprised of an integration with the navigation and editing events to track structures and update user interface elements (Section 4.1), an implementations of *fragments* (Section 4.2), and an implementation for accessing static and dynamic information of the program (Section 4.3). We provide source code references to our reference implementation in Appendix B.

### 4.1 Maintaining and Tracking Structures across Edits

To address the first challenge, HybridSE integrates two mechanisms with editor implementations: first, a structural diffing mechanism that uses heuristics to maintain the identity of program structure nodes across edits. Second, a transaction mechanism that ensures that the tool-builder-defined constraints are maintained. Together, these form the *selective structural guarantees*: HybridSE only strictly maintains structures where a tool's constraint requires it and otherwise continuously interprets textual changes for accessing information of structures.

#### 4.1.1 Diffing

Parsing program text yields a tree of program structure nodes. After the user performs a change in the text, this tree is out of date: the change may have, for example, changed the value of a number, deleted a member of an array, or introduced a syntax error that then no longer allows forming a valid tree of the program. Consequently, to find the structure of the program tree after the change, we have to parse the text again. Since there is likely a large amount of overlap between the old and the new tree, incremental parsers can reuse parse results from a previous run to accelerate a new parse [15, 30, 46]. Notably, these incremental parsing techniques do not describe how the tree was changed—they only accelerate creating a new tree. While, for example, Tree-sitter's incremental parser [30] reuses the same object identities for performance reasons, its reuse mechanism is not meant to minimize the changes in the tree and may rebuild entire subtrees even for small changes.

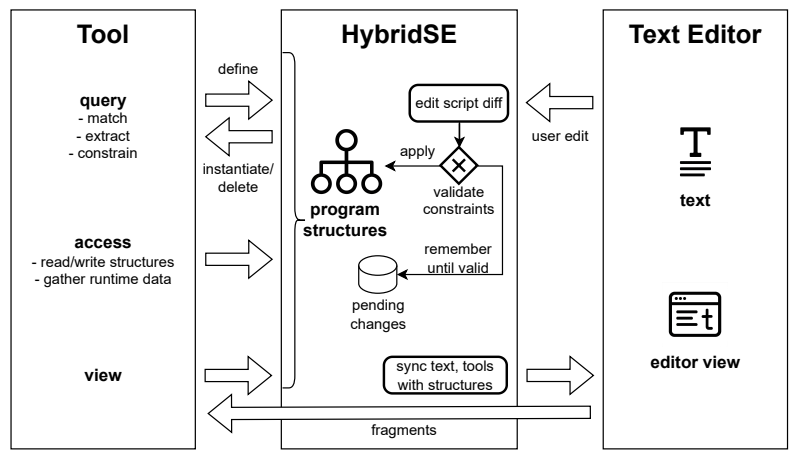

**Figure 2** Conceptual structure of HybridSE forming a bridge between tools and the host editor. HybridSE integrates with text editors but exposes structures to tools.

Since our goal is to maintain the identities of nodes such that the tools appear at the right locations and extract data from the right structures, we thus need to manually identify the implications of a change in the text for the tree structures. For this purpose, we adapted truediff [19], a structural diffing algorithm that employs a set of heuristics to compute an edit script between two versions of a program tree. The authors of truediff have evaluated the quality of their heuristics, demonstrating that it is on par with other popular structural diffing approaches. Their evaluation concerns commit-level changes, while in our use cases most diffing occurs between states that are only a few interactions apart, so results in the scenarios common to our use case are likely even better.

In general, our proposed system can be used with any structural diffing approach that produces an edit script describing the changes that have to be applied to the source tree to arrive at the target tree. Such an edit script can be made up of operations of different granularity. For example, truediff defines five basic operations:

**Load** Create a new tree node.
**Attach** Attach the given tree node to a parent.
**Detach** Detach the given tree node from its parent.
**Remove** Delete a tree node.
**Update** Update the textual content in a leaf node of the tree.

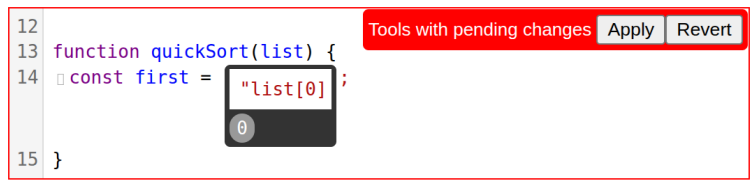

**Figure 3** The user has just inserted a quotation mark at the start of the watched expression, leading to an invalid parse tree. HybridSE displays a notice informing the user of the invalid state and offering to apply or revert their changes through the host editor.

For example, if the user types a plus sign after a number, the edit script will consist at a minimum of a detach operation for the existing number, load operations for the binary operator, the binary operation, and a placeholder node for the right hand side operand (requiring an error recovery parser), and attach operations for the operation, the operand, the operator, and the number. After calculating the edit script to arrive at the target tree, we apply it to the source tree. In this way, even though the number node changed its location in the tree in our example, it maintains its identity, as it is only detached and attached as part of the edit script.

#### 4.1.2 Transactions

The edit script that is computed by the diffing step indicates which nodes are moved, updated, or removed. Consequently, the constraints specified by our tools need only to validate that the edit script does not cause an effect that would violate them. For example, if the user adds a quotation mark in a tracked expression of our example watch defined in Listing 2, as shown in Figure 3, the watch would no longer parse correctly. To prevent this, we defined a constraint that our query should still match in the program tree that contains the user's changes in Line 9 of Listing 2.

If an edit script violates this constraint function, i.e., the constraint function returns false, the change that was made in the editor would lead to a tree that does not fulfill the expectations of an active tool. For instance, this could mean that a tool's query no longer matches, so the tool would disappear, or that a node that the tool needs to extract data from has been removed from the tree. In either case, HybridSE assumes that a violated constraint is an undesirable state that would bring the tool into an inconsistent state, as well.

Consequently, we do not apply the change to the tree. Instead, we freeze the last tree but eagerly apply the textual change in the text view of the editor and add it to a list of pending changes. At the same time, HybridSE asks the editor to communicate that an invalid state has been entered, e.g., through a highlight or a notice. Our reference implementation's notice is shown in Figure 3. HybridSE has now started a *transaction*, in which subsequent changes are collected as pending changes, and that can be resolved through one of three possible outcomes:

**Reconcile** This is the ideal scenario: the user continues editing and we eventually arrive at a tree where the constraint is no longer violated without the user performing a special intervention. After every input by the user, HybridSE creates a new edit script, going from the frozen tree to the new parse result. This behavior enables cutting-and-pasting a tool or other operations that move text with tools: when the user cuts a text range with a constraint that requires the tool to be preserved, the editor enters an invalid state. When the user pastes the text range, the diffing heuristic between the frozen tree—where the tool was at its original location—and the new tree—where the tool is at the new location—should find that the subtree with the tool was detached and then re-attached. From the tool's perspective, the intermediate state where the tool would be deleted is skipped.

**Apply** The user triggers a special editor action to apply the changes, disregarding the violated constraints. Consequently, the tool may disappear. When the user chooses to use this action, the assumption is that the user's intent is incompatible with the tool's constraints, and HybridSE prioritizes the user's intent.

**Revert** The user triggers a special editor action to revert their changes back to the last valid state, effectively an undo. Through this action, the changes in the list of pending changes are reverted, such that the text view is brought back to the state where the tool's constraints were still valid.

### 4.1.3 Host Editor Integration

To integrate HybridSE with the host editor, all textual changes performed in the host must be reported to a global `applyChanges` function shown in Listing 5. The new changes are first merged with any pending changes and then applied to the current text. This new text is then parsed and diffed against the last valid `rootNode` of the tree.

We then check if the user asked to apply the changes regardless of the tools' constraints and otherwise validate the edit script by passing it to the `validateChanges` helper method, which iterates over all constraints to check if the edit script would violate any of them. If a violation occurred, the new changes are appended to the pending changes list and the edit script is rolled back. Otherwise, if no constraints were violated given the edit script, the pending changes list can be cleared.

Finally, depending on the outcome of validation, we either update only the fragments, which we describe further in Section 4.2 or also the tools, which means to re-run their matching queries and update their views with any changed results from extraction of data of the structures.

## 4.2 Nesting Editors

The second challenge concerned nesting editors, for which we provide fragments. Fragments display the source code range of a list of consecutive program nodes. Identifying the range to be displayed is made possible through HybridSE's diffing, which maintains node identities. Tool builders can define that all or a subset of registered tool matching queries should also be active in a fragment of their tool. By matching inside a fragment, tools and fragments can nest recursively. As a tool can embed a fragment that would allow the same tool to match, there is a risk for infinite recursion, which HybridSE has a guard for.

One source of complexity for implementing fragments lies with the host editor: designing the integration of language support systems, such as autocompletion or error highlights, such that they work even when only a subrange is shown in an editor, may not be supported by default in their implementation.

The other source of complexity arises from the disjunct display of source ranges of the same source file: text navigation and selection have to consider the arrangement of fragments on the screen and the displayed text ranges. When a change to the text buffer occurred, HybridSE eventually updates the fragments, as shown in Listing 5.

The function we invoke is `updateFragments`, which first updates the text shown in each fragment. Notably, the fragment's range is derived from the nodes it shows. Consequently, when updates to the program tree are frozen, the fragment's range may have to be shifted to account for changes found in the pending changes list. The function also takes care to restore the text selection, if needed: if a shift in a fragment caused the selection to disappear or go out of sync, HybridSE attempts to restore the selection, first in the previously selected fragment, then in the fragment that is visually closest to the selected fragment that contains the requested range.

Where structure tracking forms the bridge from text to structures from the tool builder's perspective, fragments form the bridge from structures back to text from the editor user's perspective: the tool builder declares that certain structures should appear editable within the tool user interface and fragments translate that request to suit user's expectations.

Because of the transition from structure to text, fragments have to consider the secondary notation of textual layout, which is irrelevant in structured editors. This surfaces in two issues.

**Leading and Trailing Whitespace**   Expressions in textual source code commonly feature whitespace to increase legibility, for example, surrounding an infix operator like plus. As fragments track structures, they may omit whitespace that users add as they extend an expression, as described in Section 2.2. To mitigate this issue, we employ a default heuristic to always pull in whitespace to the right of a fragment's expression but only pull in whitespace to the left of a fragment's expression if it is not indentation and if it is more than one space character. Tool builders can override this behavior in case the structure they display in a fragment follows other formatting conventions, in particular depending on the language.

**Indentation**   To address the mismatch in indentation already described in Listing 1, fragments compute the largest common indentation level, omitting empty lines, and replace indentation up to that level with a single symbol tab character, as seen at the start of indented lines in Figure 1.

### 4.3 Structured Access

The third challenge formulated in Section 1 concerns reading and editing information of structures. As already described in Section 3, implementations should provide an interface to make static information about the program tree structures accessible to tools. In the following, we discuss access to dynamic information and means to edit structures.

### 4.3.1 Dynamic Information

To make dynamic information accessible, the host platform should provide an instrumentation mechanism that allows tools to establish a mapping between a reference to a program structure node and runtime values it assumes.

In our reference implementation, the mapping is facilitated through source code rewriting. Nodes for which the runtime value stream was requested, as shown in Line 6 of Listing 2, are rewritten to report values to the editor. As a simple form that works well across languages with small effort, we rewrite expressions in-place to connect to a TCP or HTTP server. The message we send includes a unique identifier for the node and a serialized form of the runtime value.

A rewritten expression for JavaScript running in the browser takes the following form in our reference implementation:

```
1 (e => (
2   fetch("https://localhost:3000/watch", {
3     method: "POST",
4     body: JSON.stringify({ id: $identifier, e }),
5     headers: { "Content-Type": "application/json" },
6   }), e)
7 )($expressions)
```

In the context of the browser, this expression can substitute the instrumented expression in place without requiring additional adaptations to the file. By wrapping the instrumentation in an immediately-evaluated function, we can bind the value of the expression such that it is evaluated only once and can be passed both to our instrumentation and returned to resume execution.

Implementations can choose to either perform rewriting to a copy of the source code or use HybridSE's tools to match against the rewritten expression and hide the instrumentation code, as we did for the marker construct in Section 3. If the rewritten expression is written to the original source code, the editor system should remove the instrumentation before code is committed to a repository or shared otherwise—writing instrumented source code to a copy is thus preferable but the choice will likely predominantly depend on the requirements of the build chain and runtime environment.

### 4.3.2 Editing Structures

With regards to editing, there is a mismatch between the program structures HybridSE passes to tool builders and the underlying textual representation. When given a program structure node, it is likely that tool authors would expect to be able to simply insert a child node in a list or move an entire subtree to another location in the tree. However, as the tree is the result of the parsing process, all changes must be performed in a manner such that a subsequent parse of the source string arrives at the same tree structure. As a consequence, edits have to take into account escaping of string values, delimiters between list elements, precedence rules and parentheses in binary expressions, or the users' formatting preferences [4].

As an example, assume that a tool wants to inline the identifier a in Python.

```
1  a = 2 + 3
2  a * 4
```

The naive text replacement approach would arrive at:

```
1  2 + 3 * 4
```

As Python defines implicit precedence rules that perform multiplication before addition, the desired source text is instead:

```
1  (2 + 3) * 4
```

Fortunately, while surface syntax often differs, most languages use similar patterns in their program trees, such that we can provide systematic support for editing.

**Atoms** For the purposes of editing, atoms are any node that can be replaced in its entirety without affecting the structure of its parents or siblings. These include literal values, such as numbers or identifiers. An array literal can also appear as an atom, as long as we are only inserting it in its entirety and not changing its contents. Similarly, a string can be considered atomic if regarded in its entirety, even if it may include interpolated identifiers, escaped characters, or special new line rules. These nested contents of a string would instead appear as the list structure described below.

**Siblings** Siblings in a tree appear where lists of structures are used. These are often separated by a delimiter string. Examples include statements separated by semicolons, arrays delimited by braces and separated by commas, or arguments to a function call separated by commas. In other lists, a separator is not needed: some items that can appear in an S-Expression (nested, parentheses-delimited lists) have well-defined extents and can thus appear directly adjacent without causing ambiguity, e.g., ((b)(c)).

**Parent-Child Structures** Frequent examples in programming languages of this type are recursive expression hierarchies, involving infix, prefix, and postfix operators. Special care needs to be taken for editing these structures, as they are subject to precedence rules.

### 4.3.3 Editing Operations

To work with these syntactic structures, HybridSE defines four editing operations:

**node.insert(string, index)** Given a string of source code, insert it in the node at the given index, adding separators as needed.

**node.delete()** Delete the node, as well as any now obsolete separators.

**node.replaceWith(string)** Replace the node with the string of source code, potentially inserting parentheses.

**node.wrapWith(string, string)** Wrap the node with the given prefix and suffix, potentially inserting parentheses.

**Hybrid Structured Editing**

While there may appear to be a mismatch between acting on structures and providing strings as arguments, we chose strings for two reasons: first, if the argument is given as a constant string, text is a concise way for authors to specify the structure they want to insert. Second, if an existing structure is given as argument instead, we may still prefer to access the string's underlying source to maintain the user's original formatting choices. Finally, the structure has to be serialized to text in either case.

To perform these operations, the first step is to access the definition in the grammar for the relevant nodes. HybridSE requires an adapter implementation for each grammar formalism to be used with editing. The adapter must implement three functions: (1) return whether a given node is in a list and what separator string that list uses, (2) return a string index for the first position to insert into a list, and (3) return whether a node can be wrapped with parentheses according to the grammar. Our reference implementation includes an adapter for Tree-sitter grammars.

To illustrate the editing operations, we consider the following simplified grammar definition and program text as a running example:

```
FunctionDeclaration := Type Identifier "(" (Type ",")* ");"
void main(int, char,);
```

**Inserting in Sibling Nodes**   Given the example above, the tool author may issue a command to insert another type as second item of the function declaration, invoking declaration.insert("float", 1). For insertion, we first check for an existing child that is in a repeating rule and find its separator string using the adapter implementation. To do so in our Tree-sitter adapter, we first use a simplified variant of partial parsing [2], which attempts to match the existing program tree nodes to the grammar. If there are multiple repeating rules, users could provide a function to select the desired one. In our example, we find the (Type ",")* rule and the int node. To identify the lists separator, we use a set of heuristics that detect typical grammatical structures for separated lists as expressed by authors of Tree-sitter grammars. Other grammar formalisms may explicitly denote delimiters or require different heuristics. Below are three examples, where A is the repeated element and "S" is the separator string:

```
at least one element, no trailing separator:
A ("S" A)*
empty list allowed, no trailing separator
(A ("S" A)*) | ϵ
empty list allowed, trailing separator allowed:
(A ("S" A)* ("S" | ϵ)) | ϵ
```

For program tree patterns that do not follow these conventions, for example, Python's tuples, which for one-element tuples appear as (1,), keeping the comma, authors can add their own heuristics or exceptions to the built-in heuristics.

We repeat this process until we reach the index that the user has requested insertion for, so in our example, we search for one more element as the user has requested insertion in the second position. Once we found the existing element that we want to insert after, we insert the given string and separator for this list. If the list is currently empty, which means we find no existing element that is in a list, we use the adapter's function to return the string index for inserting at the start of the list.

Tom Beckmann, Christoph Thiede, Jens Lincke, and Robert Hirschfeld**Deleting**   Deleting a node is only possible if the grammar allows $\epsilon$, i.e., no element, in its place or if it is part of a repeating structure. For a repeating structure, we also want to delete any separators that become obsolete after deleting the requested node. For that purpose, we employ a similar method to inserting: we first detect if the list contains separators. We then check if either of the node's adjacent siblings is such a separator, and if so, delete it as well.

**Replacing**   If the user requests to replace a node, we interpret it as a request to place the given string logically in the same position in the tree as the node to be replaced. Consequently, we may have to ensure that parentheses are set. To do so, we first ask the grammar adapter to find if it is a grammar rule that supports parentheses. If so, we insert the text that the user requested without parentheses, reparse, and try to access an program tree node at the exact range where inserted the string. If we cannot find such a node, associativity caused the parts we inserted to be assigned to different subtrees, so we repeat the replace operation with parentheses.

## 5   Reference Implementations

To demonstrate the technical feasibility of our proposed interface, we implemented the interface for CodeMirror v6 editors and drafted two implementations for Lively 4[2] and Squeak/Smalltalk [23] to identify technical hurdles. The implementation is accessible as an artifact [6].

Our reference implementation of HybridSE is for CodeMirror v6 [29], a popular browser-based editor library. CodeMirror v6 features a functional library design and is designed for extensibility. Consequently, it serves as a good starting point. As an editor library, it does not act as a full IDE, e.g., services like autocompletion are not integrated by default, so there are no challenges when it comes to displaying only a portion of a file.

As a second editor, we chose Lively 4's default editor. Lively 4 is a browser-based IDE. Its editor widget is also based on CodeMirror, but on version 5, which exposes an imperative API instead. As both CodeMirror v6 and Lively are browser-based, we were able to reuse the core implementation of HybridSE, such as the parsing and diffing algorithms, and only re-implement functionality that facilitates integration with the editor, demonstrating its potential use as an isolated, reusable API. Lively 4 integrates multiple language services for JavaScript.

Finally, we extended Squeak/Smalltalk's default editor that is implemented in Morphic [28] with HybridSE. This necessitated re-implementing HybridSE's core for Smalltalk, as well as adding extension points to Squeak's core methods, as its default editor is not designed for extensibility with regard to editing events. Squeak's text editor widget itself, however, can be configured easily to provide the expected language services for subsections of a program.

---

[2] https://github.com/LivelyKernel/lively4-core, last accessed: 2025-09-27.

1:17

**Hybrid Structured Editing**

```
1  function functionName () {
2      body
3  }
```

■ **Figure 4** Placeholder text fields for expressions or identifiers in JavaScript. When the user starts typing in a text field, it is automatically replaced with just the input text.

Beyond our reference implementation and the two implementation sketches, we began evaluating other popular editors for these capabilities. For JetBrains IDEs [24], tools that implement a similar tight integration into the source code have been demonstrated before [32]. Its API for creating extensions is sufficiently powerful to fulfill all three criteria.

Visual Studio Code's [34] extension API allows only limited access to the editor widget. Its internal API does not currently support embedding arbitrarily-sized widgets that span only a portion of a line. Instead, embedded widgets are expected to take the full horizontal space or appear as popups. Accordingly, an integration of HybridSE would require changes in Visual Studio Code's editor widget implementation.

In our reference implementation of HybridSE, we use Tree-sitter [30] for parsing. As Tree-sitter only releases parsers as executables, we access the grammar definitions from its open-source parser repositories as required for editing (Section 4.3.3). While HybridSE works with any parser, not just Tree-sitter, heuristics that derive separators or precedence have to be adapted for each new type of grammar library.

## 6 Case Studies

To evaluate the range of tools that can be created with HybridSE, we selected small tools that appear as unobtrusive additions to the text editor, as well as large-scale restructurings of file-based editing toward a more visual interface.

### 6.1 Inserting Placeholder Text in Programs

Tools will sometimes want to insert expressions that still require user input. For that purpose, HybridSE recommends defining a small tool that shows a placeholder where code is still missing, as seen in Figure 4.

The tool searches for the marker string `__VI_PLACEHOLDER_{label}`, so for example `__VI_PLACEHOLDER_body` to create a placeholder text field labeled body. When a user begins typing in the text field, it is replaced with the new input. This is a convenient means for tool builders of other tools to communicate to their users where input is still required while still producing a valid program tree without missing identifiers.

The implementation of the placeholder tool requires 36 lines of code as shown in Listing 3. This tool defines a special constraint that requires not just the node to remain present, but also for the node's content to remain the same, as the tool depends on the exact match of the node content.

```
      const db = pool.connect();
      db.query(sql`SELECT * FROM `events` WHERE id = 123`);
```

**Figure 5** A SQL editor composited within a JavaScript editor. The nested SQL is defined in a backtick-delimited string and also contains backticks itself, but no manual escaping is required, as HybridSE translates any changes with correct formatting.

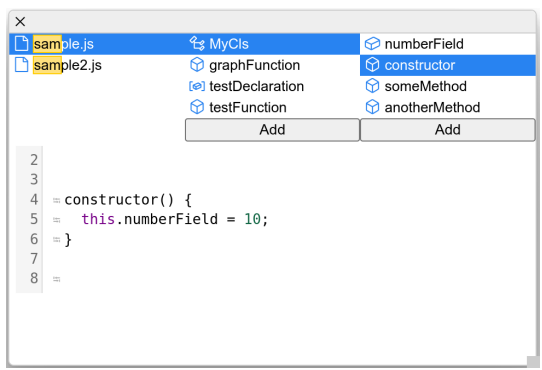

**Figure 6** A browser tool for a JavaScript class.

## 6.2 Composing Languages

Libraries commonly allow specifying source text of other programming languages as strings. One often-seen example is SQL, which is passed as a string to a database. When storing a second language in a string of a host language, users have to follow escaping rules, which may render the expression more complex. In the general case, it is not possible to compose two textual languages without escaping. This capability is one of the advantages of editors that use structures over parsing [13, 44].

As HybridSE exposes a structured interface to tools, this type of composition is well-supported. We define a tool that matches the embedded string containing the second language. When extracting the string's contents, we remove escaping and display the unescaped string in a new fragment, configured for the embedded language. In Figure 5, we show a JavaScript editor with an embedded SQL editor. Note the backticks that are used both as string delimiters in JavaScript and within SQL. When the user types a character, we check if it needs to be escaped, and if so, directly insert the escaped character in the host programming language's string.

As unescaping strings may be a common need for tools, our reference implementation offers a built-in means for JavaScript. Using that built-in, the definition of the tool comprises only 17 lines of code as shown in Listing 4. The built-in itself comprises 50 lines of code, as it also facilitates the translation of cursor indices between the escaped and unescaped versions.

## 6.3 Structural Code Browser for JavaScript

Inspired by Smalltalk's code browser [23] and drawing from CodeBubble's remapping of Java files into a per-definition structure, we constructed a browser-like interface

**Hybrid Structured Editing**

using HybridSE. A screenshot of a JavaScript file opened in the browser is shown in Figure 6. The browser interface splits a source file into individually editable declarations. This brings a number of challenges that HybridSE can solve for tool builders.

To be able to open files of a programming language in the browser, we require a mapping that defines how declarations are nested in the respective language. Optionally, the mapping may also define icons and labels for different types of declarations. For JavaScript, we have defined both icon mappings for top-level nodes, meaning every direct child of the program node, as well as a mapping that extracts nested declarations from a JavaScript class. To keep the list of top-level nodes usable, we merge some types of consecutive nodes, such as import statements.

Users open a project folder and the browser interface appears. In the left column, all project files are listed. In the middle column, the top-level statements for the selected file are shown. In the right column, any members of the selected top-level statement are shown. The panel below contains a text editor that shows only the selected program node.

The browser interface requires that individual declarations can be edited in isolation. Intermittent changes can cause preceding or following declarations to be damaged in the source tree. For example, assume a JavaScript program with two top-level statements:

```
1  var a = 5
2  b
```

If a user types a plus sign after the 5, the program will be parsed as a single statement of the form `var a = 5 + b`, despite appearing on multiple lines. As a consequence, the statement we display in the browser's editor would suddenly pull in the second statement when the plus is typed. To prevent this behavior, we define a constraint that asserts that the selected node, as well as the preceding and following nodes, remain as top-level statements, preventing the following or preceding statements from being pulled in while the user is formulating changes.

When the user clicks on the Add button in either the top-level or the member list, we call the insert method described in Section 4.3.3 with a placeholder of the form described in Section 6.1. The insert method returns the newly created symbol, allowing us to immediately select it for the edit view.

The implementation of the browser consists of the definition of its user interface as 200 lines of code, also including state management for selection. In another 160 lines of mostly query statements, the implementation expresses matching top-level statements and their nested members. These also include merging consecutive top-level items that are impractical as separate items, such as multi-line comments and the list of file imports.

### 6.4 User Interface of "Livelits"

Our final case study, shown in Figure 7, implements the user interface component of Livelits [37], fulfilling their requirements of persistence, compositionality, parametrization, and aspects of its liveness. The slider, for example, uses the `["slider", 0, 255, 1, 73][1]` form to achieve persistence and parametrization of the tool (specifying minimum, maximum, and step values). The color picker, in addition to that, accesses the runtime values of the red, green, and blue values to achieve liveness. Compositionality is achieved through HybridSE's fragments. As our example is implemented in JavaScript, we do not address the typing requirement. The decentralized extensibility requirement is also not addressed in our work—our reference implementation does bundle tool definitions as part of libraries that use the tools, however.

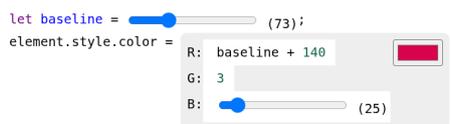

**Figure 7** Slider and color tools reimplementing Livelits [37].

## 7 Discussion and Future Work

In the following, we discuss limitations of our proposed approach and possible future work.

**Feasibility of Editor Integration** As described in Section 5, even with the proposed approach, there can still remain significant engineering effort for the host editor if it is not well-prepared for the integration of visual elements other than text. However, as open source text editors such as CodeMirror are able to offer compatible interfaces, there appears to be no conceptual limitation. There may be concerns about performance: running the matching queries over large files and instantiating large numbers of tools may degrade overall performance. Notably, tools may change the height of lines. This complicates a common optimization: when all lines have the same height, editors can calculate the precise range of lines for a given scroll position, thus skipping unnecessary text layout operations.

**Structured Access to Static and Dynamic Information** We believe HybridSE's interfaces for static and dynamic access of information demonstrates the value of an interface that uses structure node references to facilitate access. Currently, static information is limited to the syntactic level of the program structures. Instead, for some tools, it may be desirable to also access semantic information, such as references to a variable. For that purpose, a mapping between node identities and the facilities offered by language server could be created [33].

In terms of dynamic information, the demonstrated approach via source code rewriting enables a very simple means for mapping between references to structures and observed runtime values. To extend this interface further, one should consider the source of information, such as a test run, an example execution, or the actual running program. Further, the interface would benefit from more detailed information about the way execution, such as stack traces at the moment of executing a statement, ideally in a way that the stack frames can be mapped directly to program structure references in the editor.

**Hybrid Structured Editing**

**Tool Annotations in Source Code**   Many of the tools we showcased exploit the structure of the program tree to function and sometimes actively modify it for their purposes, such as the ["__watch", expr][1] marker. In some cases, these markers are considered permanent and should be checked into version control, for example the annotation that a number should appear as a slider in Section 6.4. In other cases, they may be a temporary annotation for tracking purposes, as discussing in Section 4.3.1. This approach allows tools to integrate with source code without extending the language syntax, which would render files incompatible with existing language support tools, and allows tool information to be persisted in source files such that textual version control systems can work with them.

This is in contrast to annotations in comments or in a separate file—both approaches would likely not rely on the surprising [1] that our rewriting approach requires but are likely more brittle with regards to changes in editors without HybridSE that, for instance, a colleague might use. While an editor with HybridSE that is aware of the annotations in comments or separate files can make sure to update these, users may forget to manually update them, for example, during a merge in a web interface.

Compared to other forms of annotations, the expression rewriting approach is constrained by the language syntax: for example, the method shown in our placeholder case study in Section 6.1 would not allow tools to create a dropdown for optionally adding a keyword, such as public in Java, as the keyword would have to be changed to be identified as placeholder.

Compared to structured editors that store structures instead of text, for example, editors implemented in MPS [10], our approach confronts tool builders with more challenges. Constraints do not have to be manually specified, annotations to program nodes are automatically persisted, and there is no mismatch going from the tools' structured view to the users' text view. Our proposed hybrid approach thus mostly provides value for users of the editor but also enables tool builders to leverage the likely larger ecosystem of existing tools that require text.

**Usability**   As described, our approach allows the user to continue using textual editors while also benefiting from code-centric tools. Our reference implementation has not invested significant effort into optimizing the user experience yet and future work should perform a user study to find if the proposed interactions work without flaws. Generally, the fact that interactions remained familiar from the underlying text editor allowed users in informal testing to get started using the editor without explanation beyond the specific tools' functions.

Through our own use of the reference implementation, we noticed two major issues with the way our reference implementation handles constraints: the popup that we show appears rather obtrusive. A more subtle design could show a smaller notice. Further, by only using a set of built-in constraints as opposed to the generic function that our approach proposes, it may be possible for HybridSE to derive more precisely in what way a change invalidated a constraint and could thus provide better feedback to the user, especially when constraints of multiple tools failed and caused a "freeze" of the program tree.

Second, HybridSE lacks a heuristic to recognize that the user intended to delete a tool: when the user selects a larger text range that includes a tool and hits backspace, a constraint of a tool that requested to stay in the program tree will be violated. Differentiating between backspace and the cut operation may allow HybridSE to derive user intent better.

## 8 Conclusion

Creating code-centric tools presents multiple challenges, as the text form only intermittently yields valid tree structures that tools can access. Structured editors solve this challenge but present a user interface that may be unfamiliar to users coming from text editors and also necessitate recreating large parts of a programming language ecosystem that assumes text.

We proposed *hybrid structured editing* (HybridSE), which presents tools with a fully structured interface and users with a fully textual interface. As the bridge between the two worlds, our HybridSE allows tools to selectively specify constraints on structures, and uses tree diffing and transactions between invalid states of the program tree to give tools the impression of a continuously available structured program tree. Through case studies, we demonstrated HybridSE's suitability to create code-centric tools: complexity from integrating with text buffers and accessing information from structures is shifted onto HybridSE. Hybrid structured editing thus presents a means for text editors to provide an extension mechanism for tools that support users in understanding the connection between source code and information the tools derive.

**Acknowledgements** We are grateful for the anonymous reviewers' helpful feedback. This work was supported by SAP and the HPI–MIT "Designing for Sustainability" research program.[3]

## A Code Listings

**Listing 3** Full declaration of the placeholder tool described in Section 6.1.

```
const placeholderTool = {
  query: node => {
    // mapping the grammar-specific node types for identifiers
    let identifierType = {
        "javascript": "identifier",
        "typescript": "identifier",
        "python": "identifier",
        "php": "name",
        "bash": "word",
```

---

[3] https://hpi.de/en/research/cooperations-partners/research-program-designing-for-sustainability.html, last accessed: 2025-02-04.

```
10      }[node.language];
11      if (node.type == identifierType && node.text.startsWith("__VI_PLACEHOLDER") return {};
12      else return null;
13    },
14    constraints: [
15      // prevent changes just after the identifier from changing our label
16      tool => tool.node.text === tool.originalText,
17    ],
18    view: ({ nodes, tool }) => {
19      const label = nodes[0].text
20          .substring("__VI_PLACEHOLDER_".length)
21          .replace(/_/g, " ");
22
23      tool.originalText = nodes[0].text;
24
25      return <input
26        // mark this input as representative for this source range of the file
27        ref={markInputEditableForNode(nodes[0].range)}
28        placeholder={label}
29        oninput={e =>
30          nodes[0].replaceWith(e.target.value, {
31            intentDeleteNodes: [nodes[0]],
32            requireContinueInput: true,
33          })} />
34    },
35    type: "replace",
36 };
```

■ **Listing 4** Full declaration of the sql tool described in Section 6.2.

```
1  const sqlTool = {
2    query: node => {
3      const match = node.matchesTemplate("sql`$_string`");
4      if (!match) return null;
5      // provides an unescaped string and function that translates unescaped
6      // changes back to the escaped string
7      else return { string: bindPlainString(match.string) };
8    },
9    view: ({ string: { text, onLocalChange } }) => {
10     const source = { get value() { return text; }, set value(_) {} };
11     return <CodeMirrorWithVitrail
12       onchange={(e) => e.detail.changes.forEach((c) => onLocalChange?.(c, true))}
13       cmExtensions={sql()}
14       value={source} />;
15   },
16   type: "replace",
17 };
```

■ **Listing 5** Function that informs the HybridSE of a change.

```
1  def applyChanges(changes, forceApply = false):
2    oldText = hybridSE.text
3    hybridSE.text = applyChangesToString(hybridSE.text, hybridSE.pendingChanges + changes)
4    editScript = parseAndDiff(rootNode, hybridSE.text)
```

```
5      if not forceApply and not validateChanges(editScript, changes):
6          hybridSE.text = oldText
7          updateFragments(changes)
8      else:
9          updateFragments(changes)
10         updateTools()
11
12  def validateChanges(editScript, changes):
13      if not editScript or validators.any(validator => not validator(editScript, changes.flatMap(c => c.
            ↪ nodesIntendedForDelete))):
14          editScript.rollback()
15          hybridSE.pendingChanges += changes
16          return false
17      hybridSE.pendingChanges.clear()
18      return true
```

## B   Source Code References

The following file references point to the artifact [6] (the term "vitrail" used below is the name we chose for our reference implementation of HybridSE).

- Section 4.1.1, Diffing: core/diff.js
- Section 4.1.2, Transactions: part of the applyChanges function in vitrail/vitrail.ts:532, pending changes notice implementation for CodeMirror in vitrail/codemirror6.ts:449.
- Section 4.2, Nesting Editors: vitrail/pane.ts, whitespace handling core/query.js:218, indentation handling: vitrail/tools/whitespace.ts
- Section 4.3.1, Dynamic Information: vitrail/tools/watch.ts
- Section 4.3.3, Editing Operations: core/model.js:821 and the implementation for Tree-sitter core/tree-sitter.js:337

**About the authors**

**Tom Beckmann** is a member of the Software Architecture Group of the Hasso Plattner Institute at the University of Potsdam. He is working on structured editing for general-purpose languages to better support integration of tools. His current research interests include programming tool design, as well as editing and input methods for programming. Contact Tom at tom.beckmann@hpi.uni-potsdam.de.
https://orcid.org/0000-0003-0015-1717

**Christoph Thiede** is a member of the Software Architecture Group of the Hasso Plattner Institute at the University of Potsdam. His current research interests include exploratory programming workflows, tool design, and techniques for analyzing and manipulating program execution. Contact him at christoph.thiede@hpi.uni-potsdam.de.
https://orcid.org/0000-0002-7442-8216

**Jens Lincke** is a member of the Software Architecture group at the Hasso Plattner Institute, where he is interested in live and explorative programming (Lively Kernel). He was awarded a PhD for a thesis on evolving tools in a collaborative self-supporting development environment. Contact him at jens.lincke@hpi.uni-potsdam.de.
https://orcid.org/0000-0002-3828-7778

**Robert Hirschfeld** leads the Software Architecture Group at the Hasso Plattner Institute at the University of Potsdam. His research interests include dynamic programming languages, development tools, and runtime environments to make live, exploratory programming more approachable. Contact Robert at robert.hirschfeld@hpi.uni-potsdam.de.
https://orcid.org/0000-0002-4249-6003